\documentclass{elsart4}

\usepackage{amssymb}

\usepackage{graphics}
\usepackage{graphicx}
\usepackage{epsfig}
\usepackage{amssymb}
\journal{Physica B}

\begin{document}

\begin{frontmatter}


 \title{Conductance through strongly interacting rings in a magnetic field}
 \author{Juli\'an Rinc\'on, A. A. Aligia and K. Hallberg\corauthref{cor1}}
 \address{Instituto Balseiro, Centro At\'omico Bariloche, CNEA and CONICET, 8400 Bariloche, Argentina}
 \corauth[cor1]{Tel: +54-2944-445170; e-mail: karen@cab.cnea.gov.ar}





\begin{abstract}

We study the conductance through finite Aharonov-Bohm rings of interacting electrons weakly 
coupled to non-interacting leads at two arbitrary sites.
This model can describe an array of quantum dots with a large charging energy compared to the
interdot overlap. As a consequence of the spin-charge separation, which occurs in these highly
correlated systems, the transmittance is shown to present pronounced dips for particular values
of the magnetic flux piercing the ring. We analyze this effect by numerical and analytical means
and show that the zero-temperature equilibrium conductance in fact presents these striking
features which could be observed experimentally.

\end{abstract}

\end{frontmatter}

\section{Introduction}

One of the challenges of nanotechnology is the possibility of fabricating new artificial structures
with tailored properties. For example, the Kondo effect was achieved in a system consisting of one quantum 
dot connected to leads\cite{gold1,cro,wiel}; systems of a
few QD's have been proposed theoretically as realizations of the two-channel
Kondo model~\cite{oreg,zit}, the ionic Hubbard model,~\cite{ihm}
 and the double exchange mechanism.~\cite{mart} Also, the correlation-driven metal-insulator transition 
has been studied in a chain of quantum dots.~\cite{kou} 

Another interesting phenomenon in strongly correlated systems is what is known as charge-spin separation.
It is well known that strong correlations in one dimension invalidate the Fermi liquid conventional description
of electrons. In particular, correlations can lead to the fractionalization of the electron into charge 
and spin degrees of freedom.\cite{gia} This
separation is an asymptotic low-energy property in an infinite chain. However, exact
Bethe ansatz results for the Hubbard model in the limit of infinite Coulomb
repulsion $U$ show that the wave function factorizes into a charge
and a spin part for any size of the system.~\cite{ogata} There have been several experiments reporting 
indirect indications of charge-spin separation ~\cite{Voit1,Kim,Qimiao}, and it could
also be potentially observed in systems such as cuprate chains, 
ladder compounds,~\cite{DagottoRice} and carbon nanotubes.~\cite{egger}

Several theoretical approaches tackled this phenomenon in the ring geometry. For example, the real-time evolution 
of electronic wave packets in Hubbard rings has shown a splitting in the dispersion of the spin and charge
densities as a consequence of the different charge and spin velocities.~\cite%
{Jagla1,uli} Pseudospin-charge separation has also been studied in quasi-one-dimensional quantum gases 
of fermionic atoms.~\cite{recati,ke} Other theoretical approaches concern the transmittance through 
Aharonov-Bohm rings modeled by Tomonaga-Luttinger liquid or other correlated Hamiltonians like the Hubbard 
or $t-J$ models.\cite{Jagla,meden,nos1,nos2} In these systems, noticeable dips are observed in the conductance
for fractional values of the flux which can be attributed to destructive interference between degenerate states
as we explain below. We have recently discussed the extension of these results to ladders of two legs as a
first step to higher dimensions.~\cite{julian}

In this work, we analyze the origin of the dips in the
transmittance as a function of applied flux in finite rings described by the $t-J$ model. We discuss the
conditions for which the intensity of the lowest-lying peak in the zero-temperature equilibrium
conductance as a function of the gate voltage presents characteristic dips for certain flux values. These will be
shown to be a consequence of spin-charge separation.

\section{Model Hamiltonian}

\begin{figure}
\includegraphics[width=6.5cm]{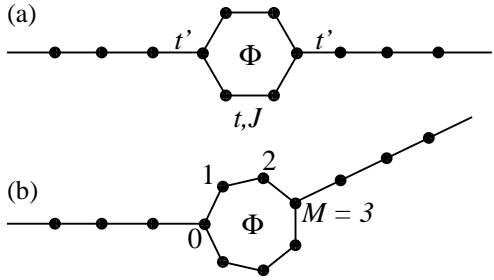}
\caption{Schematic representation of the systems studied numerically. (a) $L=6$, (b) $L=7$. In
both cases $M=3$.}
\label{dfhepta}
\end{figure}

The basic model is depicted in Fig.~\ref{dfhepta}. We have considered a ring of $L$ sites, 
weakly connected to non-interacting leads at sites $0$ and $M$. 

The Hamiltonian reads: 
\begin{equation}
H=H_{\mathrm{ring}}+H_{\mathrm{leads}}+H_{\mathrm{links}}.  \label{esh}
\end{equation}%
The first term describes the isolated ring, with on-site energy
given by a gate voltage $V_{g}$, and hoppings modified by the phase $\exp
(i\phi /L)$ due to the circulation of the vector potential. For most of the
results of this paper we use the $t-J$ model to describe the ring, 
\begin{eqnarray}
H_{\mathrm{ring}} &=&-eV_{g}\sum_{i\sigma }c_{i\sigma }^{\dag }c_{i\sigma
}-t\left( c_{i+1\sigma }^{\dag }c_{i\sigma }e^{i\phi /L}+\mathrm{H.c.}%
\right)   \nonumber \\
&&+J\sum_{i}\left(\mathbf{S}_{i}\cdot \mathbf{S}_{i+1}-\frac{1}{4}\right).  
\label{ehr}
\end{eqnarray}%
where $\phi =2\pi \Phi /\Phi _{0}$, $/\Phi _{0}=h/2e$ is the flux quantum, 
$\mathbf{S}_{i}$ is the spin operator at
site $i$ and double occupancy is not allowed at any site of the ring. The
second term corresponds to two tight-binding semi-infinite chains for the
left and right leads 
\begin{equation}
H_{\mathrm{leads}}\!=\!-t\!\!\!\sum_{i=-\infty ,\sigma
}^{-1}\!\!\!a_{i-1,\sigma }^{\dag }a_{i,\sigma }\!-t\!\!\!\sum_{i=1,\sigma
}^{\infty }\!a_{i,\sigma }^{\dag }a_{i+1,\sigma }\!+\!\mathrm{H.c.}
\label{ehl1}
\end{equation}
The third term in Eq.~(\ref{esh}) describes the coupling of the left (right) lead with
site 0 ($M$) of the ring 
\begin{equation}
H_{\mathrm{links}}=-t^{\prime }\sum_{\sigma }(a_{-1,\sigma }^{\dag
}c_{0\sigma }+a_{1,\sigma }^{\dag }c_{M\sigma }+\mathrm{H.c.}).  \label{ehl2}
\end{equation}

\section{Conductance}

When the ground state of the isolated ring is non-degenerate, and the
coupling $t^{\prime }$ between the leads and ring is weak, the equilibrium
conductance at zero temperature can be expressed to second order in $%
t^{\prime }$ in terms of the retarded Green's function for the isolated ring
between sites $i$ and $j$: $G_{i,j}^{\mathrm{R}}(\omega )$.~\cite{ihm,Jagla}
For an incident particle with energy $\omega =-2t\cos k$ and momentum 
$\pm k$, the transmittance reads 
\begin{equation}
T(\omega ,V_{g},\phi )=\frac{4t^{2}\sin ^{2}k|{\tilde{t}}(\omega )|^{2}}{%
\left\vert [\omega -{\epsilon }(\omega )+te^{ik}]^{2}-|{\tilde{t}}%
^{2}(\omega )|\right\vert ^{2}},  \label{tra}
\end{equation}%
where $\epsilon (\omega )=t^{\prime \,2}G_{00}^{\mathrm{R}}(\omega )$ and  
$\tilde{t}(\omega )=t^{\prime \,2}G_{0M}^{\mathrm{R}}(\omega )$
represent a correction to the on-site energy at the extremes of the
leads and an effective hopping between them respectively.

This equation is in fact exact for a non-interacting system, however, it loses
validity for an odd number of electrons, where the ground state is Kramers degenerate. 
In this case the method misses completely the interesting physics arising from the
Kondo effect. \cite{ihm,lobos} However, the Kondo temperature is small compared to the other 
energy scales of the system and we can assume safely that the Kondo effect is 
destroyed by a small temperature. 
This approach is justified for small enough $t^{\prime }$ since the characteristic energy 
of this Kondo effect decreases exponentially.

The conductance is $G=(ne^{2}/h)T(\mu ,V_{g},\phi )$, where $n=1$ or 2
depending if the spin degeneracy is broken or not,~\cite{ihm} and $\mu $ is
the Fermi level, which we set as zero (half-filled leads). When the gate
voltage $V_{g}$ is varied a peak in the conductance is obtained when there
is a degeneracy in the ground state of the ring for two consecutive number
of particles: $E_{g}(N+1)=E_{g}(N)$, where $E_{g}(N)$ is the ground state
energy of $H_{\mathrm{ring}}$ with $N$ electrons. Without loss of generality, 
we assume that we start with $N+1 $ electrons in the ring and apply a negative gate voltage in such a way
that a peak in the conductance is obtained at a critical value $V_{g}^{c}$
when the number of electrons in the ring changes from $N+1$ to $N$
electrons.

\section{Analytical and numerical results}

For $J=0$ the model is equivalent to the Hubbard model with infinite
on-site repulsion $U$, for which the wave function can be factorized into a
spin and a charge part, evidencing charge-spin separation.~\cite{ogata,nos1,caspers}
For each spin state, the system can be mapped
into a spinless model with an effective flux which depends on the total
spin. For a system of $N$ particles one can construct
spin-wave functions with wave vectors
$k_{s}=2\pi n_{s}/N$, where the integer $n_{s}$ characterizes the spin wave
function. The total energy and momentum (in an appropriate gauge) of any
state of the ring have simple expressions: 
\begin{eqnarray}
E &=&-2t\sum_{l=1}^{N}\cos (k_{l}),\;\;k_{l}=\frac{2\pi n_{l} +\phi _{%
\mathrm{eff}}}{L},  \label{ene} \\
K &=&\sum k_{l}=\left[ 2\pi (n_{c}+n_{s})+N\phi \right] /L,  \label{k} \\
\phi _{\mathrm{eff}}&=&\phi +k_{s}=\phi +\frac{2\pi }{N}n_{s},
\label{phieff}
\end{eqnarray}
where the integers $n_{l}$ characterize the charge part of the wave function
and $n_{c}=\sum n_{l}$.

The values of the flux $\phi _{d}$ for which dips or reduced conductances
are expected, correspond to some particular crossings of the energy levels
of $N$ electrons. One can see this from the general form of the Green's functions
$G_{0j}^{\mathrm{R}}(\omega )$ entering the 
transmittance [Eq.~(\ref{tra})] when a particle is destroyed.
Using the Lehman's representation, the relevant part of the Green's function is: 
\begin{equation}
G_{0j}^{\mathrm{R}}(\omega )=\sum_{e}\frac{\langle g|c_{j\sigma }^{\dagger
}|e\rangle \langle e|c_{0\sigma }|g\rangle }{\omega +E_{e}-E_{g}}.
\label{g1}
\end{equation}
Noting that the ring has translational symmetry and naming 
$K_{\nu }$ the wave vector of the state $|\nu \rangle $, one obtains:

\begin{equation}
G_{0j}^{\mathrm{R}}(\omega )=\sum_{e}\frac{e^{-ij(K_{e}-K_{g})}|\langle
e|c_{0\sigma }|g\rangle |^{2}}{\omega +E_{e}-E_{g}}.  \label{gr}
\end{equation}

At certain flux values, two states of $N$ electrons, $|e\rangle $
and $|e^{\prime }\rangle $, become degenerate. Assuming that the corresponding
matrix elements entering Eq.~(\ref{gr}) are nonzero, only these two
states, contribute significantly to the Green's function $G_{0j}^{\mathrm{R}}(\omega )$ at
the Fermi energy ($\omega =\mu =0$) when $V_{g}$ (which displaces all $E_{e}$
rigidly with respect to $E_{g}$) is tuned in such a way that $%
E_{e}=E_{e^{\prime }}\sim E_{g}$. Denoting by $\beta =\exp [iM(K_{e^{\prime }}-K_{e})]$ 
the relative phase between the two intervening degenerate states we see that if 
$\beta \neq 1$, the transmittance, which is proportional to $|G_{0M}^{\mathrm{R}}(\omega )|^{2}$, 
(Eq.~(\ref{tra})), is reduced near the crossing (note that this results is independent of any 
specific model).

We can predict the positions of the dips in the transmittance without having to resort to the 
calculation of the matrix elements entering the Green´s functions  Eq.~(\ref{gr}). The crossings of energy
levels at low energies take place at $\phi =-\pi (n_{s}+n_{s}^{\prime })/N$.\cite{nos2} When 
$n_{s}+n_{s}^{\prime } $ is odd (even) the relative phase $\beta =\exp
[iL(K_{e^{\prime }}-K_{e})/2]=\exp \left[ i(n_{s}^{\prime }-n_{s})\right] $
\ $=-1$ (1) and there is (there is not) a dip in the integrated
transmittance. Therefore, the positions of the dips are located at 
\begin{equation}
\phi _{d}=\pi (2n+1)/N,  \label{dip}
\end{equation}
with $n$ integer. These are also the positions where crossings in the
(experimentally accessible) ground state for $N$ particles take place ($%
n_{s}^{\prime }-n_{s}=\pm 1$).

\begin{figure}
\includegraphics[width=7.5cm]{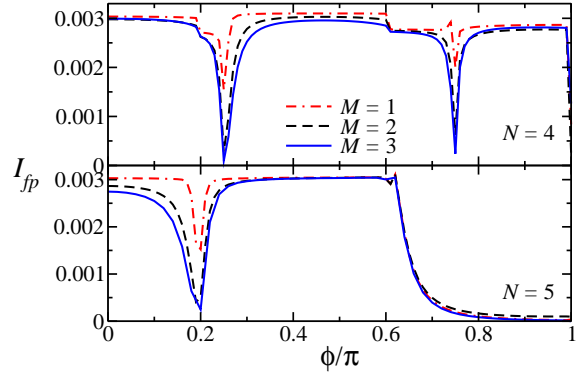}
\caption{Intensity of the first peak in the
transmittance $I_{fp}$, as a function of applied
magnetic flux, for a ring of 6 sites and  (a) $N+1=5$ and (b) $N+1=6$ electrons in the ground state,  
three different configurations $M=1$, 2 and 3, $t^{\prime}=0.3t$ and $J=0.001t$.}
\label{Fig2}
\end{figure}

To check these predictions we have performed numerical calculations for the transmittance,
obtained by diagonalizing the ring using Davidson's method~\cite{david}. Once the Green's
functions were obtained, they were replaced in Eq.~(\ref{tra}), to obtain the transmittance. The
systems studied are represented in Fig.~\ref{dfhepta}. In contrast to previous
work,~\cite{Jagla,meden,nos1,julian} we concentrate on the first peak in the transmittance as
the gate voltage is decreased since this is the feature which is experimentally accessible at 
equilibrium and low temperatures.

In Fig.~\ref{Fig2} we show numerical results for a ring with $L=6$ sites and $N=4$ and 5 particles 
in the intermediate state for the three non-equivalent configurations corresponding to $M=1$, 2 and 3.
To quantify the relative intensity of the conductance we integrate the
transmittance given by Eq.~(\ref{tra}) over a window of gate
voltage $V_{g}$ of width $0.002t$ centered around the degeneracy point
between the ground state for $N+1$ and $N$ electrons. This corresponds to
the intensity of the first observable peak in the transmittance as the
gate voltage is lowered.  As the curve is symmetric under change of sign of $\phi$
we show only the interval $0\leq \phi \leq \pi $. For $%
J\rightarrow 0$ and $N=4$ the dips appear as expected (Eq.~(\ref{dip})) at  $\phi _{d}/\pi =0.25$ and
0.75. For $N=5$ 
the dips should occur at $\phi _{d}/\pi =0.2$, 0.6 and 1. However, near 0.6 the total spin
in the ground state for 5 electrons changes from $S=1/2$ to $S=3/2$, which is not 
accessible by destroying an electron in the 6-electron singlet ground
state. Therefore, the transmittance vanishes for $0.6<\phi
\leq \pi $. As a consequence, in the interval shown there is only one dip
present. 
In this figure one can see the effect of the different source-drain configurations: the lowest conductance 
is achieved for the symmetric case ($M=3$ for $L=6$) where $\beta=-1$ and both degenerate levels interfere 
destructively. For the other cases the interference is less destructive and a less pronounced dip is obtained.

\begin{figure}
\vspace{0.3cm}
\includegraphics[width=7.5cm]{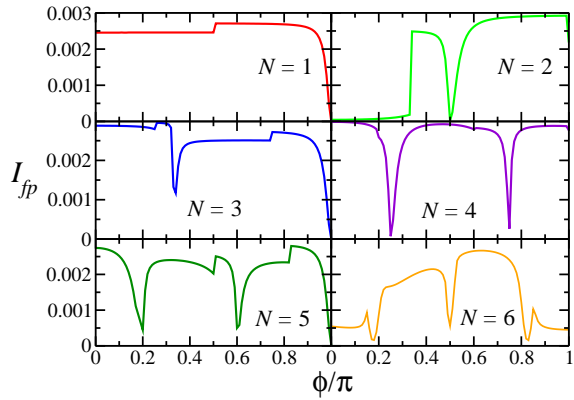}
\caption{Intensity of the first peak in the
transmittance $I_{fp}$, as a function of applied
magnetic flux, for a ring of 7 sites and several fillings. Here $M=3$, $t^{\prime}=0.3t$ and 
$J=0.001t$.}
\label{Fig3}
\end{figure}

It is also interesting to analyze the conductance for different particle numbers to study the evolution of
the dips as predicted in Eq.~(\ref{dip}). This is shown in Fig.~\ref{Fig3}. 
The number of particles in the intermediate situation, $N$, is shown for each case. Here we see that the minima 
do in fact occur for the fluxes predicted by that relation very accurately. The abrupt step obtained for $N=2$ 
corresponds, again, to a forbidden transition to a large total spin state from a singlet ground 
state as explained above.

\section{Conclusions}

We have presented results for the zero-temperature \emph{equilibrium} conductance
through finite rings described by the $t-J$ model threaded by a magnetic
flux, weakly coupled to conducting leads. At particular values of the flux we find dips or reductions 
of the transmittance, which are due to negative interferences between degenerate levels.
This can be understood by analyzing the extremely interacting case for $J=0$, where exact results are
available. The position of the dips reflect the particular features
of the spectrum in this limit, in which the charge and spin degrees of freedom
are separated at all energies. For finite $J$ the positions of the dips change and some additional dips can
also appear in a manner that is difficult to predict and which is not yet fully understood.

The negative interference depends on the source-drain configuration. 
It is more marked if the leads are connected at angles near 180 degrees. These results are confirmed by our
numerical calculations. 

\section*{Acknowledgments}

This investigation was sponsored by PIP 5254 of CONICET and PICT 2006/483 of
the ANPCyT. 



\end{document}